\documentclass[prb,twocolumn,superscriptaddress,showpacs,preprintnumbers]{revtex4}
\usepackage{amssymb}

\usepackage{graphicx}
\usepackage{dcolumn}
\usepackage{bm}

\begin{document}

\title{Insulating behavior with spin and charge order in the ionic
  Hubbard model} 

\author{Krzysztof~Byczuk}
 \affiliation{Institute of Theoretical Physics, 
University of Warsaw, ul. Ho\.za 69, PL-00-681 Warszawa, Poland}
\author{Michael~Sekania}
\affiliation{Theoretical Physics III, Center for Electronic
  Correlations and Magnetism,
Institute of Physics, University of Augsburg, D-86135 Augsburg,
Germany}
\author{Walter~Hofstetter}
\affiliation{Institut f\"ur Theoretische Physik, Johann Wolfgang
  Goethe-Universit\"at,
D-60438 Frankfurt/Main, Germany}
\author{Arno~P.~Kampf}
\affiliation{Theoretical Physics III, Center for Electronic Correlations and 
Magnetism, Institute of Physics, University of Augsburg, D-86135 Augsburg,
Germany}

\date{\today }

\begin{abstract}
Paramagnetic solutions of the ionic Hubbard model at half-filling in 
dimensions $D>2$ indicate that the band and the Mott insulator phases are 
separated by a metallic phase. We present zero-temperature dynamical 
mean-field theory solutions, which include antiferromagnetic long-range order, 
and show that the one-particle spectral functions always possess an energy gap 
and therefore the system is insulating for all interaction strengths. The 
staggered charge density modulation coexists with antiferromagnetic long-range
order of N\'eel type.  
\end{abstract}

\pacs{71.27.+a, 71.30.+h, 71.10.Fd}
\maketitle

A bipartite lattice system of non-interacting electrons with one particle per 
site is a perfect gapless metal. Applying an external alternating potential 
with a periodicity of twice the lattice constant $a$ doubles the unit cell, 
thereby reduces the Brillouin zone (BZ), and opens a gap at the BZ boundary. 
Such a system is a \emph{band insulator} with a charge density modulation with 
wavelength $2a$; this system is also referred to as an \emph{ionic insulator}. 
\cite{Gebhard97} On the other hand, switching on a local repulsive 
interaction between the electrons with opposite spins leads either to a 
paramagnetic \emph{Mott-Hubbard insulator} with a correlation induced energy 
gap or to the spontaneous development of antiferromagnetic (AF) long range 
order. In the latter case it is the presence of AF order which doubles the
unit cell and opens a gap at the BZ boundary for weak interactions and thereby 
creates a \emph{Slater insulator}. In the strong interaction limit the 
electrons are localized with antiferromagnetically aligned spins forming a
\emph{Mott-Heisenberg insulator}. Experimental and theoretical investigations 
of metal-insulator transitions and transitions between different insulators 
continue as a challenge for condensed matter physics.  

The \emph{ionic Hubbard model} \cite{Hubbard81,Nagaosa86} incorporates both 
interactions and an external alternating potential and is therefore well 
suited to study transitions between metallic or different insulating phases. 
This model was originally used to study the neutral-ionic transition in 
organic charge transfer salts \cite{Nagaosa86} or ferroelectric transitions in 
perovskite materials.\cite{Egami93} But the understanding of possible phase 
transitions in the ionic Hubbard model may prove important for other strongly 
correlated electron systems as well such as, for example, FeSi.\cite{Kunes08} 
The physics of the ionic Hubbard model may even find a realization in optical 
lattices, if two laser beams of commensurate wavelengths are superposed with 
properly tuned amplitudes.\cite{Grainer08}

Extensive literature records exact, approximate, as well as numerical results 
for the ionic Hubbard model in one dimension.\cite{Gogolin,Kampf03,Manmana04} 
It is agreed that at half-filling and in the interaction dominated regime the 
system is a paramagnetic Mott insulator whereas in the alternating potential 
dominated regime the system is an ionic band insulator. By now there is an 
emergent consensus that these two types of insulators are separated by yet
another insulating phase with a non-zero \emph{bond-order} parameter, which 
is the expectation value for a staggered component of the kinetic energy. 

More recently the ionic Hubbard model was also investigated in higher 
dimensions within single-site or cluster dynamical mean-field theory (DMFT) 
\cite{Jabben05,Garg06,Craco07,Kancharla07} or by determinant quantum 
Monte-Carlo simulations.\cite{Paris07} Using DMFT with iterated perturbation 
theory as the tool for solving the DMFT equations, Garg \emph{et al.} 
determined the ground-state phase diagram of the ionic Hubbard model at 
half-filling for a semicircular density of states.\cite{Garg06}  With the
restriction to paramagnetic solutions an intermediate metallic phase was found
separating the Mott and the band insulator. Within the same computational 
framework Craco \emph{et al.} identified a coexisting phase between two 
insulators as well as discontinuous metal-insulator transitions.\cite{Craco07}
The discontinuous transitions were confirmed in the two dimensional (2D) 
system by Kancharla \emph{et al.}, \cite{Kancharla07} who used cluster DMFT 
combined with exact diagonalization and interpreted their data in favor of an 
intermediate bond ordered phase. 

While the DMFT work was restricted to paramagnetic solutions, finite 
temperature quantum Monte Carlo in 2D simulations also probed AF
correlations.\cite{Paris07}  The presence of the intermediate metallic
phase between the  
band and the Mott insulator was confirmed. However, since in the 2D system at 
finite temperature long range antiferromagnetism is
prohibited,\cite{Mermin} the question how the possible presence of the
antiferromagnetic long-range order in higher dimensions or in $2D$ at
zero temperature changes the phase diagram has remained open.

Here we apply DMFT to the ionic Hubbard model at zero temperature allowing 
for spontaneous AF long-range order. We find that the ground state is
always insulating, with a gap in the one-particle spectral function.
There is a direct transition between the band insulator and the AF
Mott insulator. Beyond a critical interaction strength 
the charge- and the spin-density modulations coexist. In this region the 
insulator has AF character. 

The ionic Hubbard model on a bipartite lattice is defined by the following
Hamiltonian
\begin{eqnarray}
H=-t\sum_{\langle ij\rangle\sigma} a^{\dagger}_{i\sigma}a^{}_{j\sigma}+
U\sum_i n_{i\uparrow}n_{i\downarrow}+\sum_{i\sigma}\Delta_i n_{i\sigma}\, ,
\label{hamiltonian}
\end{eqnarray}
where $\Delta_i=\pm \Delta/2$ for $i \in A$, $B$ sublattices, respectively.
The first term describes the kinetic energy of the electrons with spin 
$\sigma=\pm 1/2$ for the hopping between nearest-neighbor lattice sites $i$ 
and $j$ with amplitude $t$, the Hubbard interaction leads to an energy 
increase $U\geq 0$ for the double occupancy of a site, and the last term 
contains a staggered potential with an energy difference $\Delta\geq 0$ 
between the $A$ and $B$ sublattices. The operators $a^{}_{i\sigma}$ and 
$a^{\dagger}_{i\sigma}$ obey standard fermionic anticommutation relations and 
$n_{i\sigma}=a^{\dagger}_{i\sigma}a^{}_{i\sigma}$.

The Hamiltonian (\ref{hamiltonian}) is solved within DMFT at half-filling, 
i.e. for the chemical potential $\mu=U/2$, by a mapping to two 
non-equivalent impurity problems with the ionic energies $\pm
\Delta/2$.\cite{Metzner89,Vollhardt93,Pruschke95,Georges96}  The
impurity sites are  
coupled to two particle baths. Explicitly, we solve separately the two 
different single-impurity Anderson models
\begin{eqnarray}
H_{\rm SIAM}^{\alpha}&=&(\epsilon_{\alpha}-\mu)n_{\alpha\sigma}
+\sum_{k}V_{k\alpha \sigma}\left(a^{\dagger}_{\alpha\sigma} c_{k\alpha\sigma} 
+ h.c.\right) \nonumber \\
&+&Un_{\alpha\uparrow}n_{\alpha\downarrow}
+ \sum_k
\epsilon_{k\alpha \sigma} 
c^{\dagger}_{k\alpha\sigma} c^{}_{k\alpha\sigma} ,
\label{siam}
\end{eqnarray}
where the hybridization matrix elements $V_{k\alpha \sigma}$ and the kinetic 
energies of the bath electrons $\epsilon_{k\alpha\sigma}$ are obtained 
self-consistently by additional DMFT equations.\cite{Georges96} Here we 
explicitly keep the spin and the site ($\alpha=A,B$) dependences in order to 
allow selectively for spin and charge order. The one-particle impurity energy 
is $\epsilon_{\alpha}=\pm\Delta/2$ for $\alpha=A$ or $B$, respectively. The 
single-impurity Anderson Hamiltonians are solved at zero temperature by 
the numerical renormalization group (NRG) method.\cite{Bulla08} This method 
allows to obtain the spectral functions at and near the Fermi energy with high
precision and can therefore accurately distinguish between metallic and 
insulating phases. \cite{Bulla99}  

The local (impurity) Green functions obtained from (\ref{siam}) are expressed 
via the hybridization function $\eta_{\alpha\sigma}(\omega)$ and the 
self-energies $\Sigma_{\alpha\sigma}(\omega)$ as
\begin{eqnarray}
G_{\alpha\sigma}(\omega)=\frac{1}{\omega
  -(\epsilon_{\alpha}-\mu)-\eta_{\alpha \sigma}(\omega) -
  \Sigma_{\alpha\sigma}(\omega)} \, .
\label{green}
\end{eqnarray}
The hybridization functions describe the resonant broadening of the impurity 
energy levels due to the coupling to the particle baths and are given by 
\begin{equation}
\eta_{\alpha\sigma}(\omega)=\sum_k \frac{V_{k\alpha
    \sigma}}{\omega-\epsilon_{k\alpha\sigma}} .
\label{hybridization}
\end{equation}
The self-energies capture the interaction induced correlation effects on the
impurity sites. 

\begin{figure}[t]
\centering
\includegraphics[width=0.99\columnwidth]{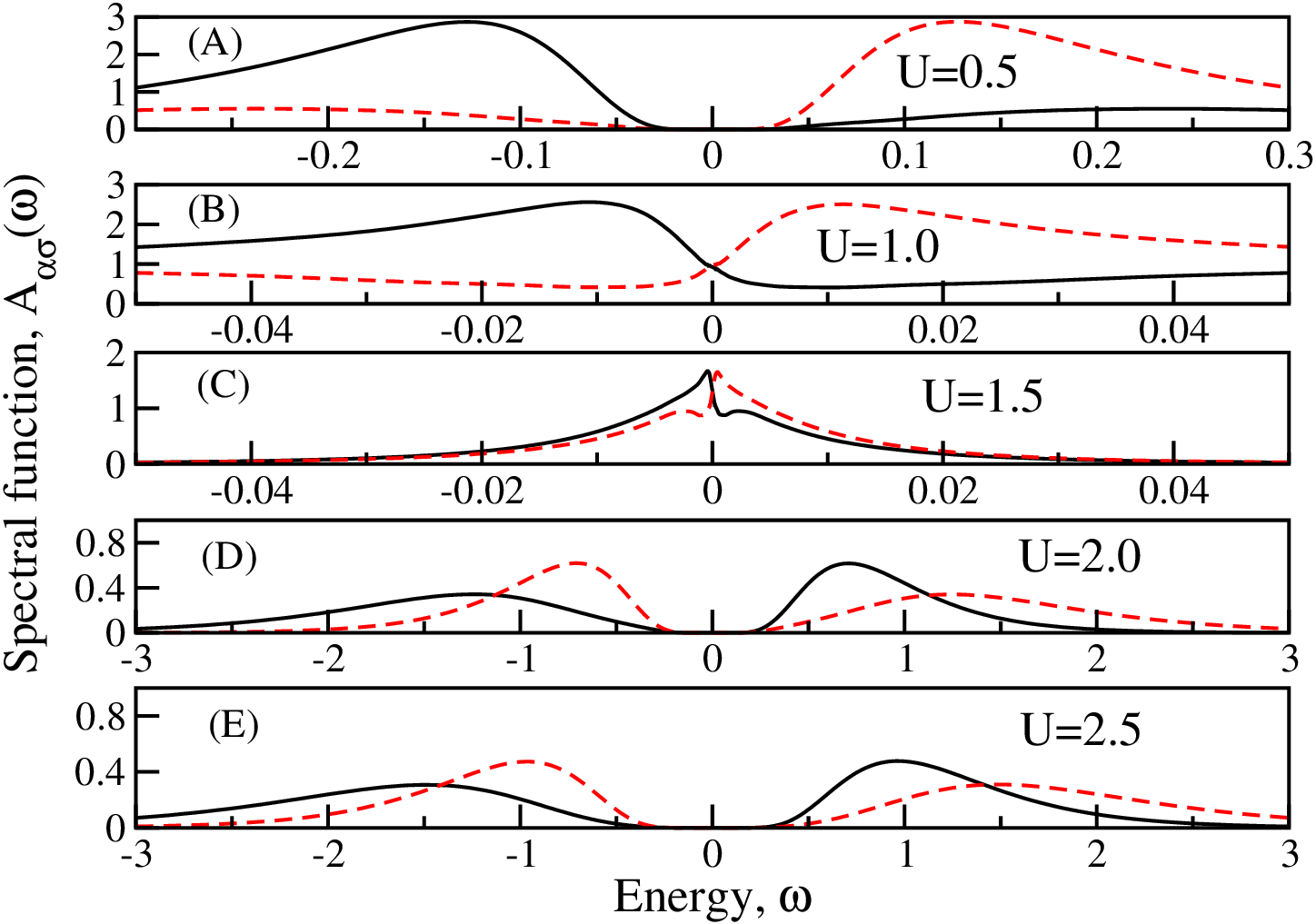}
\caption{(color online) 
Spectral functions for the ionic Hubbard model at $\Delta=0.5$ and 
different interactions $U$ in the paramagnetic limit. Solid and dashed lines 
correspond to the A and B sublattices. From top to bottom: (A) band 
insulator; (B), (C) correlated metal; (D), (F) correlated Mott insulators. 
Note the different scales on the axis in particular horizontal ones. }
\label{fig1}
\end{figure}

Within DMFT the hybridization functions are subject to self-consistency 
conditions, which involve the density of states (DOS) for a given lattice 
structure.\cite{Vollhardt93,Pruschke95,Georges96} In the following we adopt 
the semicircular DOS corresponding to the Bethe
lattice.\cite{Georges96,Eckstein05,Kollar05} 
 The hybridization functions are then 
simply related to the local Green functions (\ref{green}) through
\begin{equation}
\eta_{\alpha\sigma}(\omega)=\frac{W^2}{16}G_{\bar{\alpha}\bar{\sigma}}(\omega),
\label{selfconsistency}
\end{equation}
where $\bar{\alpha}=B,A$ if $\alpha=A,B$ and $\bar{\sigma}=-\sigma$, 
respectively. $W=1$ is the bandwidth, which sets the energy unit. We emphasize
that the use of the semicircular DOS has merely technical reasons because this
choice simplifies the DMFT equations. The obtained results remain 
qualitatively similar for any particle-hole symmetric DOS, which represents a 
bipartite lattice in dimensions $D>2$.\cite{Eckstein05,Kollar05}

From the self-consistent DMFT solution of the ionic Hubbard model we determine
(i) the local one-particle spectral function $A_{\alpha\sigma}(\omega)=-
{\rm Im}\, G_{\alpha\sigma}(\omega)/\pi$, (ii) the AF order parameter 
(staggered magnetization) $m_{\rm AF}=\langle n_{A\uparrow}-n_{A\downarrow}
\rangle=-\langle n_{B\uparrow}-n_{B\downarrow}\rangle$, and (iii) the charge 
density wave amplitude (ionicity) $m_{\rm CDW}=\langle n_{A\uparrow}+
n_{A\downarrow}-n_{B\uparrow}-n_{B\downarrow}\rangle$, where $\langle\cdots
\rangle$ denotes ground-state expectation values. A metal is distinguished 
from an insulator by a finite spectral function at the Fermi level, i.e. 
$A(0)=\sum_{\alpha\sigma}A_{\alpha\sigma}(0)>0$. 

\begin{figure}[t]
\includegraphics[width=0.99\columnwidth]{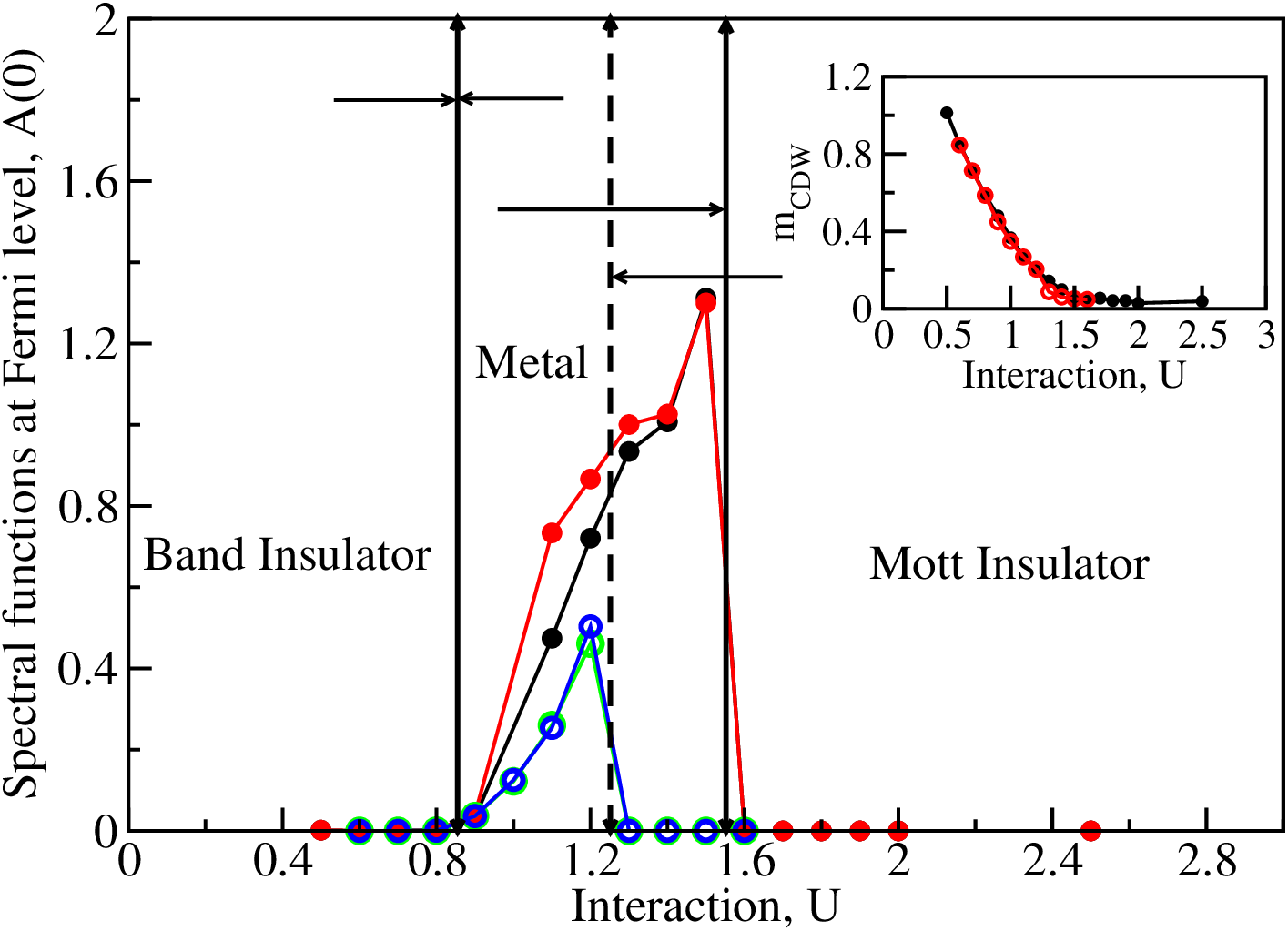}
\caption{(color online) 
Spectral functions at the Fermi energy versus $U$ for the ionic 
Hubbard model at $\Delta=0.5$ in the paramagnetic limit. Solid data points
(black and red) are obtained from DMFT iterations which start from an initial
metallic input whereas open points (blue and green) are obtained from
an insulating  input. The band insulator to metal transition is
continuous, but the metal to  Mott insulator transition is
hysteretic. Inset: charge density wave amplitude $m_{\rm CDW}$
vs. $U$.  
}
\label{fig1_bis}
\end{figure}

Without allowing for long-range AF order in the self-consistent DMFT 
solution the results of iterated perturbation theory suggested that the band 
and the Mott insulator are separated by a metallic
phase.\cite{Garg06,Craco07} 
Here, we confirm this conclusion by using NRG for solving the DMFT equations. 
In Fig.~\ref{fig1} we show spectral functions in the paramagnetic regime at 
fixed $\Delta=0.5$ starting from a band insulator at small interactions (cf. 
top panel with $U=0.5$). By increasing $U$ we find a continuous transition to 
a metallic solution with finite spectral weight at the Fermi energy (for $U=1$
and $1.5$ in Fig.~\ref{fig1}B and C). This metallic phase coexists with long-range 
charge order, i.e. $m_{\rm CDW}>0$. Larger interaction strengths homogenize 
the system and the charge density wave amplitude continuously decreases (cf. 
the results for $U=1.5$ in Fig.~\ref{fig1}C and the inset in 
Fig.~\ref{fig1_bis}). A further increase of $U$ leads to a Mott-Hubbard type 
metal-insulator transition with hysteretic behavior.\cite{Bulla99,Bulla01} In 
the examples shown in Fig.~\ref{fig1} for $U=2$ and $2.5$ we find  Mott 
insulators with correlation induced spectral gaps. In the Mott insulator the 
charge density wave is very small but remains finite.

The two transitions are separated in the phase diagram and imply the existence 
of two critical interaction strengths.\cite{Garg06,Craco07} The transition 
from the metal to the Mott insulator resembles the one, which is found within 
DMFT applied to the paramagnetic Hubbard model in the absence of a staggered 
potential.\cite{Bulla99,Bulla01} At zero temperature this transition is 
continuous, though hysteretic behavior in the iterative solution is 
encountered; it occurs at $U_c\approx 1.45$ for a semicircular DOS. 

We note that although $m_{\rm CDW}$ becomes vanishingly small for large $U$ 
(see the inset in Fig.~\ref{fig1_bis}) it is expected to remain finite for all 
interaction strengths as long as $\Delta>0$. This expectation relies on the 
general argument that the symmetry of the ground state cannot be higher than 
the symmetry of the Hamiltonian itself.

It is understood,  however, that the paramagnetic solution of the Hubbard 
model at half-filling is not the generic case. Antiferromagnetism is likely 
to occur unless it is prohibited e.g. by strong frustration effects. In the 
next step we therefore discuss the ionic Hubbard model at half-filling on a 
bipartite lattice allowing for long-range antiferromagnetism.

\begin{figure}[t]
\centering
\includegraphics[width=0.99\columnwidth]{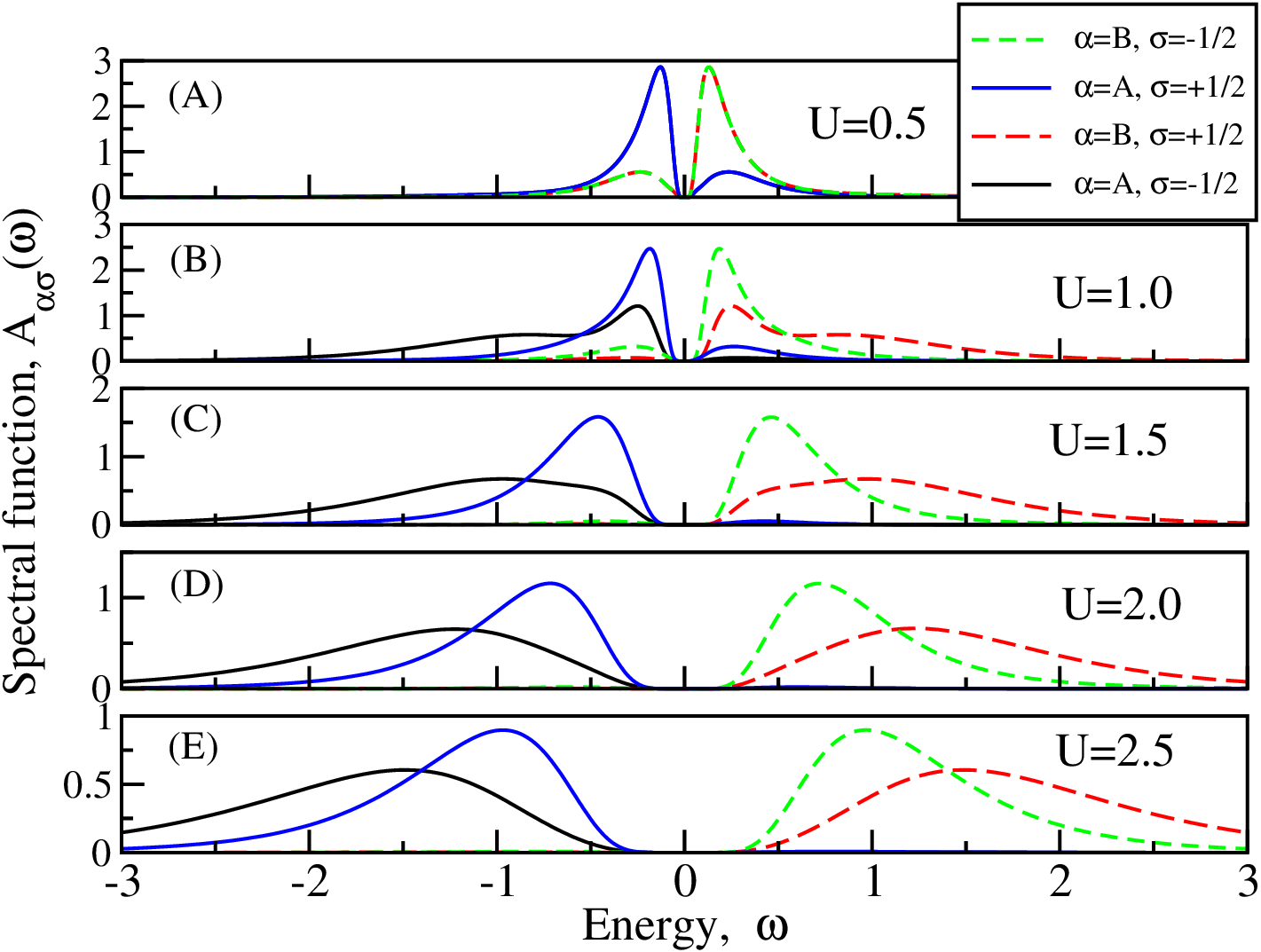}
\caption{(color online)
Spectral functions of the ionic Hubbard model for $\Delta=0.5$ and
different interactions $U$ allowing for AF order. Solid and dashed lines 
correspond to different sublattices, the color code distinguishes the spins 
$\sigma=\pm 1/2$, as indicated in the inset. (A) Paramagnetic band insulator 
with $m_{\rm CDW}>0$ and $m_{\rm AF}=0$. (B) -- (E) Antiferromagnetic ionic 
insulator with $m_{\rm CDW}>0$ and $m_{\rm AF}>0$. Note different scales on 
vertical axis. In all cases the spectra are gapped.
}
\label{fig2}
\end{figure}

Both, staggered charge order, induced by the alternating potential $\Delta$, 
and spontaneous staggered AF order may give rise to a gapped spectrum and 
insulating behavior of the lattice system at half-filling. If both staggered 
orders develop simultaneously, the above mentioned metallic phase in between 
the ionic band and the Mott insulator may be insulating due to an AF induced 
energy gap. Indeed, as the examples in Fig.~\ref{fig2} illustrate, the 
spectral functions are gapped around the Fermi energy in the entire $U$ range.
At weak $U$ the system possesses charge order only; staggered magnetization is
zero (cf. Fig.~\ref{fig2}A). By increasing $U$ $m_{CDW}$ is reduced and AF 
correlations develop. The local interaction $U$ reduces the
amount of double occupancy -- needed to maintain a finite $m_{CDW}$ at 
half-filling. On the other hand, virtual hopping processes induce an effective 
exchange interaction, which favors antiferromagnetism. Beyond a critical  $U(\Delta)$ 
the system acquires N\'eel order (cf. 
Fig.~\ref{fig2} B - E) accompanied by a strong reduction of the CDW amplitude. 

\begin{figure}[t]
\centering
\includegraphics[width=0.99\columnwidth]{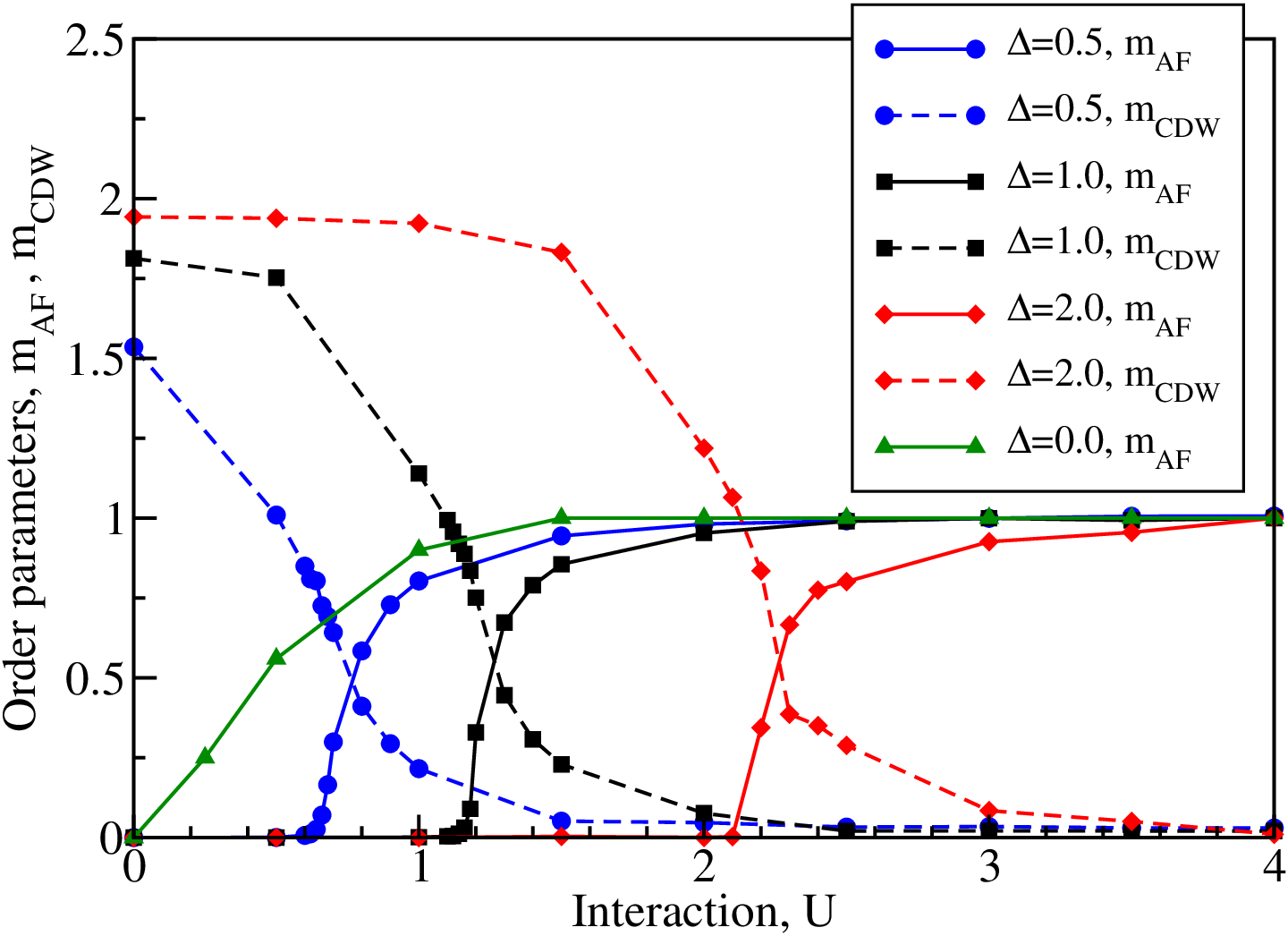}
\caption{(color online)
$U$ dependence of spin- and charge-density wave order parameters
 for different ionic potentials $\Delta$. 
}
\label{fig3}
\end{figure}

In Fig.~\ref{fig3} we present how the order parameters $m_{\rm CDW}$ and 
$m_{\rm AF}$ vary with the interaction $U$ for different ionic potentials 
$\Delta$. As discussed above, due to the suppression of double occupancies by 
the on-site repulsion, the charge density wave order parameter is reduced but 
$m_{\rm CDW}>0$ for all $U$ as inferred from symmetry arguments. Whereas at 
$\Delta=0$ N\'eel order appears at infinitesimally small $U$, at finite 
$\Delta$ the interaction has to exceed a finite critical value $U_c(\Delta)$ 
for the onset of antiferromagnetism. This quantum phase transition at 
$U=U_c(\Delta)$ is continuous in contrast to the paramagnetic-antiferromagnetic
transition for models with frustration.\cite{Zitzler04}  At very large $U$ 
the AF order parameter saturates at its maximum value.

In summary, our numerical solution of the DMFT equations for the ionic Hubbard
model provides evidence for the existence of a critical interaction 
strength for the transition from a weakly correlated band insulator to a Mott 
insulator with coexisting charge and staggered spin order. A gap in the 
one-particle spectrum persists in all parameter regimes and thus 
implies the absence of an intervening metallic phase. We emphasize that our 
results do not contradict the findings in 
Refs.~\onlinecite{Garg06,Craco07,Kancharla07,Paris07} because in those works 
the AF long-range order was either excluded by the choice of the method or by 
the low dimensionality of the system and finite temperatures. Transitions 
between a correlated metal and a band insulator are also found in other 
models, such as the Hubbard model with binary alloy disorder 
\cite{Byczuk04,Balzer04} or the bilayer Hubbard model with interlayer
hopping.\cite{Fuhrmann06}  As we have demonstrated here, the occurrence of spontaneous 
staggered long-range order can significantly change the nature or even the 
existence of a transition to a metallic phase.

We thank D. Vollhardt for insightful discussions. This work was 
supported by the Sonderforschungsbereich 484 of the Deutsche 
Forschungsgemeinschaft (DFG). The main part of the calculations were
performed when K.B. was at Augsburg University.


\end{document}